\documentclass[aps,floats]{revtex4}
\usepackage{amsmath,amssymb}
\usepackage{graphicx,epsfig}
\usepackage[greek,english]{babel}

\begin{document}
\bibliographystyle {plain}

\def\oppropto{\mathop{\propto}} 
\def\opsimeq{\mathop{\simeq}}
\def\opoverderline{\mathop{\overline}}
\def\operarrow{\mathop{\longrightarrow}}
\def\opsim{\mathop{\sim}}

\def\fig#1#2{\includegraphics[height=#1]{#2}}
\def\figx#1#2{\includegraphics[width=#1]{#2}}


\title{ Real-space renormalization 
for the finite temperature statics and dynamics \\ 
of the Dyson Long-Ranged Ferromagnetic and Spin-Glass models  } 


\author{ C\'ecile Monthus }
 \affiliation{Institut de Physique Th\'{e}orique, 
Universit\'e Paris Saclay, CNRS, CEA,
91191 Gif-sur-Yvette, France}

\begin{abstract}
The finite temperature dynamics of the Dyson hierarchical classical spins models is studied via real-space renormalization rules concerning the couplings and the relaxation times. For the ferromagnetic model involving Long-Ranged coupling $J(r) \propto r^{-1-\sigma}$ in the region $1/2<\sigma<1$ where there exists a non-mean-field-like thermal Ferromagnetic-Paramagnetic transition, the RG flows are explicitly solved: the characteristic relaxation time $\tau(L)$ follows the critical power-law $\tau(L)\propto L^{z_c(\sigma)} $ at the phase transition and the activated law $\ln \tau(L)\propto L^{\psi} $ with $\psi=1-\sigma$ in the ferromagnetic phase. For the Spin-Glass model involving random Long-Ranged couplings of variance $\overline{J^2(r)} \propto r^{-2\sigma}$ in the region $2/3<\sigma<1$ where there exists a non-mean-field-like thermal SpinGlass-Paramagnetic transition, the coupled RG flows of the couplings and of the relaxation times are studied numerically : the relaxation time $\tau(L)$ follows some power-law $\tau(L)\propto L^{z_c(\sigma)} $ at criticality and the activated law $\ln \tau(L)\propto L^{\psi} $ in the Spin-Glass phase with the dynamical exponent $\psi=1-\sigma=\theta$ coinciding with the droplet exponent governing the flow of the couplings $J(L) \propto L^{\theta} $.

\end{abstract}

\maketitle

\section{ Introduction  }

The statistical physics of equilibrium is based 
on Boltzmann's ergodic principle, that states the equivalence
between the 'time average' of any observable $A$ over a sufficiently long time $t$
and an 'ensemble average' over microscopic
configurations $C$ of energies $U(c)$
\begin{eqnarray}
\frac{1}{t} \int_0^{t} d\tau A(\tau) \opsimeq_{t \to \infty}
 \sum_{ {\cal C}} A \left(  {\cal C}  \right) 
p_{eq} \left( {\cal C}   \right)
\label{ergodic}
\end{eqnarray} 
where 
\begin{eqnarray}
P_{eq}({\cal C}) = \frac{ e^{- \beta U({\cal C})} }{Z}
\label{peq}
\end{eqnarray}
represents the Boltzmann distribution
at inverse temperature $\beta$ with the corresponding partition function
\begin{eqnarray}
Z = \sum_{\cal C}  e^{- \beta U({\cal C})}
\label{partition}
\end{eqnarray}
Even if historically and physically, this dynamical interpretation
of the equilibrium is essential, it is sometimes somewhat 'forgotten'.
In particular to discuss the appearance of low temperature
symmetry broken phases in pure systems like ferromagnets, 
it has become usual to reason only {\it statically} in terms of the properties
 of the Boltzmann measure in the thermodynamic limit, because
the possible 'pure states' are completely obvious.
For disordered systems, many discussions
of the equilibrium are also based on the same purely 'static' point of view,
but they face the huge problem
that whenever disorder induces some frustration as in spin-glasses,
the possible 'pure states' are not at all obvious 
(see for instance the book \cite{SGbook} and references therein).
For such systems, a much clearer physical
description can be thus achieved by returning to the 
'historical' point of view of statistical mechanics
where the equilibrium is considered as the
stationary measure of some {\it dynamics},
so that the question on the number of 'phases' 
for the equilibrium becomes
a question of ergodicity-breaking for the dynamics \cite{Palmer,Palmer83}.
This broken-ergodicity point of view is actually 
also the 'historical' point of view for the spin-glass problem :
in their original paper \cite{EA}, 
Edwards and Anderson have defined their order parameter by
the following sentence :
`` if on one observation a particular spin is $S_i(0)$,
then if it is studied again a long time later, there is a non-vanishing
probability that $S_i(t)$ will point in the same direction''.
The importance of this definition has been further emphasized 
by Anderson in \cite{And_houches} : ``If the spins are going to polarize
in a particular random function [...], we had better not try to characterize
the order by some kind of long-ranged order in space,
or by some kind of order parameter defined in space,
but  { \it we must approach it from a pure
non-ergodic point of view, as a long-range order in the time alone } :
 if the system has a certain order at $t=0$, then as $t \to + \infty$ 
there remain a finite memory of that order''.

 In the present paper, we thus wish to detect the presence of the low-temperature ordered
phase via the renormalization of the stochastic dynamics at finite temperature. The relevant valleys in configuration space are thus obtained as the longest-lived valleys of the dynamics.
This idea has been used previously to define a renormalization procedure 
of the master equation {\it in configuration space} \cite{us_valleys}. 
Here we focus on the Dyson hierarchical spin models in order to derive explicit renormalization rules {\it in real space}.

The paper is organized as follows. 
In section \ref{sec_dynquant}, we recall how the large-time properties
of the finite-temperature stochastic dynamics of classical Ising models
are related to the low-energy states of related quantum spin Hamiltonians.
In section \ref{sec_rg}, we introduce a real-space renormalization procedure
for the case of Dyson hierarchical Ising models.
In section \ref{sec_ferro}, the RG flows for the Dyson hierarchical
ferromagnetic model involving Long-Ranged couplings
$J(r) \propto r^{-1-\sigma}$ are explicitly computed in the region $0<\sigma<1$ where there exists a thermal Ferromagnetic-Paramagnetic transition.
In section \ref{sec_sg}, we consider the RG flows for the Dyson hierarchical
Spin Glass model involving random Long-Ranged couplings of variance $\overline{J^2(r)} \propto r^{-2\sigma}$ in the region $1/2<\sigma<1$ where there exists a thermal SpinGlass-Paramagnetic transition. Our conclusions are summarized in \ref{sec_conclusion}.

\section{ Reminder on the stochastic dynamics of classical Ising models  }

\label{sec_dynquant}

\subsection{ Finite-temperature stochastic dynamics satisfying detailed balance  }

Let us consider a generic Ising model of $N$
classical spins $S_i=\pm 1$ defined by the energy function
for the configurations ${\cal C}=(S_1,S_2,...,S_N) $
\begin{eqnarray}
 U({\cal C})  = - \sum_{1 \leq i <j \leq N} J_{ij} S_i S_j 
\label{Uener}
\end{eqnarray}
The stochastic relaxational dynamics towards the Boltzmann equilibrium 
of Eq. \ref{peq}
can be described by some master equation for the
probability $P_t ({\cal C} ) $ to be in configuration ${\cal C}$
 at time t
\begin{eqnarray}
\frac{ dP_t \left({\cal C} \right) }{dt}
= \sum_{\cal C '} P_t \left({\cal C}' \right) 
W \left({\cal C}' \to  {\cal C}  \right) 
 -  P_t \left({\cal C} \right) 
\left[  \sum_{ {\cal C} '} W \left({\cal C} \to  {\cal C}' \right) \right]
\label{master}
\end{eqnarray}
where the transition rates $W$ satisfy the detailed balance property
\begin{eqnarray}
e^{- \beta U({\cal C})}   W \left( \cal C \to \cal C '  \right)
= e^{- \beta U({\cal C '})}   W \left( \cal C' \to \cal C   \right)
\label{detailed}
\end{eqnarray}

In this paper, to simplify the notations,
 we will focus on the following single-spin-flip dynamics :
 the configuration $ {\cal C}=(S_1,S_2,...,S_N) $
containing $N$ spins is connected to the $N$ configurations
 $ {\cal C}_k=(S_1,S_2,.,-S_k,..,S_N) $ 
obtained by the flip of the single spin $S_k \to -S_k$
with the rate
\begin{eqnarray}
W \left( {\cal C} \to {\cal C}_k \right)
= \frac{1}{\tau_k }   e^{- \frac{\beta}{2} \left[ U({\cal C }_k)-U({\cal C }) \right] } 
= \frac{1}{\tau_k } e^{- \beta S_k \left(\sum_{i \ne k} J_{ij} S_i    \right)}
\label{Wsimple}
\end{eqnarray}
where $\tau_k$ is the characteristic time to attempt the flip of the spin $k$.
Although the initial dynamics usually corresponds to a uniform $\tau_k=\tau$,
the real-space renormalization introduced later
 will require to allow some $k$-dependence in the flipping times.

\subsection{ Associated quantum Hamiltonian }

As is well known (see for instance the textbooks \cite{gardiner,vankampen,risken}),
 the non-symmetric master Eq.  \ref{master}
can be transformed via the change of variable
\begin{eqnarray}
P_t ( {\cal C} ) \equiv e^{-  \frac{\beta}{2} U(\cal C ) } \psi_t ({\cal C} )
=   e^{-  \frac{\beta}{2} U(\cal C ) } <{\cal C} \vert  \psi_t  >
\label{relationPpsi}
\end{eqnarray}
into the imaginary-time Schr\"odinger equation
for the ket  $\vert  \psi_t  >$ 
\begin{eqnarray}
\frac{ d }{dt} \vert  \psi_t  > = -H \vert  \psi_t  > 
\label{Hquantum}
\end{eqnarray}
where the quantum Hamiltonian reads in terms of Pauli matrices
for the choice of the transition rates of Eq. \ref{Wsimple}
\begin{eqnarray}
H = \sum_{k=1 }^N  \frac{1}{\tau_k} \left[ e^{- \beta \sigma_k^z \left(\sum_{i \ne k} J_{ij}\sigma_i^z    \right)}   -\sigma_k^x \right]
\label{tight}
\end{eqnarray}
This type of quantum mapping for the stochastic dynamics
 has been much used both for pure spin models
\cite{glauber,felderhof,siggia,kimball,peschel,us_rgdyndyson}
and for disordered spin models 
\cite{us_conjugate,castelnovo,us_rgdyngen,us_rgdyntree,us_rgdynfractal,us_conjugateLRSG,us_LRSGfull}.

The quantum Hamiltonian $\cal H$ has the following well-known properties

(i) the ground state energy is 
\begin{eqnarray}
E_0=0
\label{e0}
\end{eqnarray}
 and the corresponding
eigenvector reads
\begin{eqnarray}
\vert  \psi_0 > = \sum_{\cal C}  \frac{ e^{- \frac{\beta}{2} U({\cal C}) }}{\sqrt Z}
\vert { \cal C}  >
\label{psi0}
\end{eqnarray}
where the normalization $1/\sqrt Z$ comes from the quantum normalization of 
eigenfunctions.
This property ensures the convergence towards the Boltzmann equilibrium
 in Eq. \ref{relationPpsi} for any initial condition ${\cal C}_0$.

(ii) the other $(2^N-1)$ 
energies $E_n>0$ determine the whole spectrum of relaxation times towards equilibrium
via exponential factors $e^{- E_n t}$. So the lowest energies
of the quantum Hamiltonian correspond to the inverses of
the largest relaxation times of the stochastic dynamics.
As a consequence, the properties of the large-time dynamics can be studied via the 
renormalization of the quantum Hamiltonian : this idea has been applied
already for various pure and disordered spin models {\it in the limit of very low
temperature $T \to 0$ for the stochastic dynamics, corresponding to very high inverse temperature $\beta=1/T \to +\infty$ } \cite{us_rgdyndyson,us_rgdyngen,us_rgdyntree,us_rgdynfractal,us_LRSGfull}. The aim of the present paper is introduce a renormalization
approach for any finite temperature $T$ of the dynamics, in order to study not only the low-temperature ordered phase (either ferromagnetic or spin-glass), but also the
critical dynamics at the phase transition towards the paramagnetic phase.

\section{ Renormalization approach for the stochastic dynamics  }

\label{sec_rg}

\subsection{  Dyson hierarchical classical spin models }

In the following, we will consider both 
the Dyson hierarchical ferromagnetic Ising model
\cite{dyson,bleher,gallavotti,book,jona,baker,mcguire,Kim,Kim77,dysonhopfield,us_rgdyndyson}
and the Dyson hierarchical spin-glass model
 \cite{franz,castel_etal,castel_parisi,castel,angelini,guerra,barbieri,us_LRSGground,us_LRSGfull}. The energy function of these Dyson hierarchical classical models
for $N=2^n$ spins is defined as a sum over 
the contributions over the generations $k=0,1,..,n-1$
\begin{eqnarray}
U_{(1,2^n)} &&  = \sum_{k=0}^{n-1} U^{(k)}_{(1,2^n)} 
\label{recDyson}
\end{eqnarray}
The generation $k=0$ contains the lowest order couplings $J^{(0)}_{2i-1,2i}$ 
\begin{eqnarray}
 U^{(k=0)}_{(1,2^n)} &&  =
 - \sum_{i=1}^{2^{n-1}}  J^{(0)}_{2i-1,2i}   S_{2i-1} \sigma_{2i}^x
\label{h0dyson}
\end{eqnarray}
the next generation $k=1$ reads
\begin{eqnarray}
 U^{(k=1)}_{(1,2^n)} &&  =
 - \sum_{i=1}^{2^{n-2}}  
\left[ J^{(1)} _{4i-3,4i-1}   S_{4i-3} S_{4i-1}
+  J^{(1)} _{4i-3,4i}   S_{4i-3} S_{4i}
+ J^{(1)} _{4i-2,4i-1}   S_{4i-2} S_{4i-1}
+  J^{(1)} _{4i-2,4i}   S_{4i-2} S_{4i}
 \right]
\label{h1dyson}
\end{eqnarray}
and so on up to the last generation $k=n-1$ that couples 
all pairs of spins between the two halves of the system
\begin{eqnarray}
 U^{(n-1)}_{(1,2^n)} = - \sum_{i=1}^{2^{n-1}} \sum_{j=2^{n-1}+1}^{2^n} 
 J^{(n-1)}_{i,j}  S_i   S_j 
\label{hlastdyson}
\end{eqnarray}
The ferromagnetic case and the spin-glass case will be more precisely defined 
by properties of the couplings $ J^{(k)}_{i,j} $ in the corresponding sections \ref{sec_ferro} and \ref{sec_sg}.

\subsection{ Quantum Hamiltonian associated to the dynamics of a block of two spins }

For the block of two spins $(S_{2i-1},S_{2i})$ 
coupled via the zero-generation couplings $J_{2i-1,2i}^{(0)}$,
the internal energy function reads (Eq. \ref{h1dyson})
\begin{eqnarray}
U(S_{2i-1},S_{2i}) &&  = - J_{2i-1,2i}^{(0)} S_{2i-1}S_{2i}
\label{u1dyson}
\end{eqnarray}
so that the quantum Hamiltonian of Eq. \ref{tight} associated to the dynamics
\begin{eqnarray}
H_{2i-1,2i} = \left(\frac{1}{\tau_{2i-1} }  + \frac{1}{\tau_{2i} } \right) 
 e^{ -\beta J^{(0)}_{2i-1,2i} \sigma_{2i-1}^z  \sigma_{2i}^z }
   - \frac{1}{\tau_{2i-1}}\sigma_{2i-1}^x 
   - \frac{1}{\tau_{2i}}\sigma_{2i}^x 
\label{tightising12}
\end{eqnarray}
acts in the $z$-basis $\vert S_{2i-1},S_{2i}>$ as
\begin{eqnarray}
H_{2i-1,2i} \vert S_{2i-1},S_{2i}> = e^{  - \beta J^{(0)}_{2i-1,2i} S_{1} S_{2}} \left(\frac{1}{\tau_{2i-1} }
  + \frac{1}{\tau_{2i} } \right) \vert S_{2i-1},S_{2i}>- \frac{1}{\tau_{2i-1} }  \vert -S_{2i-1},S_{2i}>- \frac{1}{\tau_{2i} }  \vert S_{2i-1},-S_{2i}>
\label{hquantum4}
\end{eqnarray}

\subsection{ Diagonalization in the symmetric sector }

In the symmetric sector
\begin{eqnarray}
 \vert u_S> && = \frac{ \vert ++ >+ \vert -- >}{\sqrt{2}}
\nonumber \\
 \vert v_S> && =\frac{ \vert +- >+ \vert -+ >}{\sqrt{2}}
\label{uvs}
\end{eqnarray}
Eq \ref{hquantum4} yields
\begin{eqnarray}
H_{2i-1,2i}  \vert u_S> && = e^{  - \beta J_{2i-1,2i} } \left(\frac{1}{\tau_{2i-1} }
  + \frac{1}{\tau_{2i} } \right)\vert u_S> -  \left(\frac{1}{\tau_{2i-1} }
  + \frac{1}{\tau_{2i} } \right)\vert v_S>
\nonumber \\
H_{2i-1,2i}  \vert v_S> && = -  \left(\frac{1}{\tau_{2i-1} }
  + \frac{1}{\tau_{2i} } \right)\vert u_S> + e^{   \beta J_{2i-1,2i} } \left(\frac{1}{\tau_{2i-1} }
  + \frac{1}{\tau_{2i} } \right)\vert v_S>
\label{huvs}
\end{eqnarray}
The lowest eigenvalue is of course vanishing (Eq. \ref{e0})
\begin{eqnarray}
\lambda_{2i}^{S-} =0
\label{lambdasm}
\end{eqnarray}
and the associated eigenvector corresponds to Eq \ref{psi0}
\begin{eqnarray}
\vert \lambda_{2i}^{S-} > = \frac{e^{  \frac{ \beta}{2} J^{(0)}_{2i-1,2i} } \vert u_S>+e^{ - \frac{ \beta}{2} J_{2i-1,2i} } \vert v_S>  }{\sqrt{ 2 \cosh \beta J^{(0)}_{2i-1,2i}} }
=  \frac{e^{  \frac{ \beta}{2} J^{(0)}_{2i-1,2i} } (\vert ++ >+ \vert -- >)
+e^{ - \frac{ \beta}{2} J^{(0)}_{2i-1,2i} } (\vert +- >+ \vert -+ >)  }
{ 2 \sqrt{  \cosh \beta J^{(0)}_{2i-1,2i}} }
\label{lambdasmeigen}
\end{eqnarray}
The other eigenvalue
\begin{eqnarray}
\lambda_{2i}^{S+} =  \left(\frac{1}{\tau_{2i-1} }
  + \frac{1}{\tau_{2i} } \right) 2 \cosh \beta J_{2i-1,2i}^{(0)}
\label{lambdasp}
\end{eqnarray}
  will be later projected out.

\subsection{ Diagonalization in the antisymmetric sector }

In the antisymmetric sector
\begin{eqnarray}
 \vert u_A> && = \frac{ \vert ++ >- \vert -- >}{\sqrt{2}}
\nonumber \\
 \vert v_A> && =\frac{ \vert +- >- \vert -+ >}{\sqrt{2}}
\label{uvsa}
\end{eqnarray}
 Eq \ref{hquantum4} yields
\begin{eqnarray}
H_{2i-1,2i}  \vert u_A> && =  e^{  - \beta J^{(0)}_{2i-1,2i} } \left(\frac{1}{\tau_{2i-1} }
  + \frac{1}{\tau_{2i} } \right)\vert u_A> + \left(\frac{1}{\tau_{2i-1} }
  - \frac{1}{\tau_{2i} } \right)\vert v_A>
\nonumber \\
H_{2i-1,2i}  \vert v_A> && = \left(\frac{1}{\tau_{2i-1} }
  - \frac{1}{\tau_{2i} } \right)\vert u_A>+ e^{   \beta J^{(0)}_{2i-1,2i} } \left(\frac{1}{\tau_{2i-1} }
  + \frac{1}{\tau_{2i} } \right)\vert v_A>
\label{huva}
\end{eqnarray}
The two eigenvalues read
\begin{eqnarray}
\lambda_{2i}^{A\pm} =
 \left(\frac{1}{\tau_{2i-1} }  + \frac{1}{\tau_{2i} } \right) 
\left[  \cosh \beta J^{(0)}_{2i-1,2i} 
\pm
 \sqrt{   \sinh^2 \beta J^{(0)}_{2i-1,2i} 
+  \left(\frac{\tau_{2i}- \tau_{2i-1}}{\tau_{2i}+\tau_{2i-1} }  \right)^2 }
\right]
\label{eigena}
\end{eqnarray}
In the following we will keep the lowest 
eigenvalue $\lambda_{2i}^{A-} $ corresponding to the eigenvector
\begin{eqnarray}
\vert \lambda_{2i}^{A-}> =  u_{2i} \vert u_A> + v_{2i} \vert v_A>
\label{eigenav}
\end{eqnarray}
with the coefficients
\begin{eqnarray}
u_{2i} && = \sqrt{ \frac{1}{2} \left[1+ \frac{  \sinh \beta J^{(0)}_{2i-1,2i}}
{\sqrt{   \sinh^2 \beta J^{(0)}_{2i-1,2i} 
+  \left(\frac{\tau_{2i-1}   - \tau_{2i}}{\tau_{2i-1}   + \tau_{2i}}  \right)^2 }} \right] }
\nonumber \\
v_{2i} && = {\rm sgn} (\tau_{2i-1}-\tau_{2i}) \sqrt{ \frac{1}{2} \left[1- \frac{  \sinh \beta J^{(0)}_{2i-1,2i}}
{\sqrt{   \sinh^2 \beta J^{(0)}_{2i-1,2i} 
+  \left(\frac{\tau_{2i-1}   - \tau_{2i}}{\tau_{2i-1}   + \tau_{2i}}  \right)^2 }} \right] }
\label{UV}
\end{eqnarray}

\subsection{ Projection onto the two lowest eigenvalues }

We wish to keep the two lowest eigenvalues $\lambda_{2i}^{S-}=0 $ and $\lambda_{2i}^{A-} $.
The corresponding eigenstates are labelled
with some renormalized spin $\sigma_{R(2i)}$
\begin{eqnarray}
\vert \sigma_{R(2i)}^x =+ >  &&  \equiv \vert \lambda_{2i}^{S-} > 
\nonumber \\
\vert \sigma_{R(2i)}^x =- >  &&  \equiv  \vert \lambda_{2i}^{A-} > 
\label{spinRx}
\end{eqnarray}
or equivalently
\begin{eqnarray}
\vert \sigma_{R(2i)}^z =+ >  &&
  \equiv \frac{\vert \lambda_{2i}^{S-} >  +\vert \lambda_{2i}^{A-} > }{\sqrt{2}}
\nonumber \\
\vert \sigma_{R(2i)}^z =- >  &&
  \equiv \frac{\vert \lambda_{2i}^{S-} >  - \vert \lambda_{2i}^{A-} > }{\sqrt{2}}
\label{spinRz}
\end{eqnarray}

It is convenient to introduce the corresponding spin operators
\begin{eqnarray}
 \sigma_{R(2i)}^z   &&  \equiv \vert \sigma_{R(2i)}^z =+ > < \sigma_{R(2i)}^z =+ \vert
-  \vert \sigma_{R(2i)}^z =- > < \sigma_{R(2i)}^z =- \vert
 =   \vert \lambda_{2i}^{S-} > <\lambda_{2i}^{A-} \vert +\vert \lambda_{2i}^{A-} > <\lambda_{2i}^{S-} \vert  
\nonumber \\
 \sigma_{R(2i)}^x   &&  \equiv \vert \sigma_{R(2i)}^z =+ > < \sigma_{R(2i)}^z =- \vert
+  \vert \sigma_{R(2i)}^z =- > < \sigma_{R(2i)}^z =+ \vert
= \vert  \lambda_{2i}^{S-} > <\lambda_{2i}^{S-} \vert -  \vert \lambda_{2i}^{A-} > <\lambda_{2i}^{A-} \vert  
\label{spinRop}
\end{eqnarray}
as well as the projector
\begin{eqnarray}
P^-_{2i} \equiv \vert \lambda_{2i}^{S-} > <\lambda_{2i}^{S-} \vert +\vert \lambda_{2i}^{A-} > <\lambda_{2i}^{A-} \vert  
\label{proj}
\end{eqnarray}

\subsection{ Renormalization rule for the flipping time }

By construction, the projection of the quantum Hamiltonian 
of Eq. \ref{tightising12} reads with $\lambda_{2i}^{S-}=0 $ (Eq. \ref{lambdasm})
\begin{eqnarray}
P^-_{2i}{\cal H}_{2i-1,2i} P^-_{2i} && =  \lambda_{2i}^{S-} \vert \lambda_{2i}^{S-} > <\lambda_{2i}^{S-} \vert 
+\lambda_{2i}^{A-} \vert \lambda_{2i}^{A-} > <\lambda_{2i}^{A-} \vert
= \lambda_{2i}^{A-} \vert \lambda_{2i}^{A-} > <\lambda_{2i}^{A-} \vert  
\nonumber \\
&&= \frac{ \lambda_{2i}^{A-} }{2} \left[ P^-_{2i} -  \sigma_{R(2i)}^x  \right]
\label{projH}
\end{eqnarray}
so that the renormalized flipping time $\tau_{R(2i)}$ of the renormalized spin $\sigma_{R(2i)}$ reads
\begin{eqnarray}
\tau_{R(2i)} &&= \frac{2} { \lambda_{2i}^{A-}   }
= \frac{2} { \left(\frac{1}{\tau_{2i-1} }  + \frac{1}{\tau_{2i} } \right)  \cosh \beta J^{(0)}_{2i-1,2i} 
- \sqrt{ \left(\frac{1}{\tau_{2i-1} }  + \frac{1}{\tau_{2i} } \right)^2  \sinh^2 \beta J^{(0)}_{2i-1,2i} 
+  \left(\frac{1}{\tau_{2i-1} }  - \frac{1}{\tau_{2i} } \right)^2 } }
\nonumber \\
&& = \frac{ \left(\tau_{2i-1} +\tau_{2i}  \right)}{ 2 } 
\left[  \cosh \beta J^{(0)}_{2i-1,2i} 
+ \sqrt{   \sinh^2 \beta J^{(0)}_{2i-1,2i} 
+  \left(
\frac{\tau_{2i-1}   - \tau_{2i}}{\tau_{2i-1}   + \tau_{2i}}
 \right)^2 } \right] 
\label{tauR}
\end{eqnarray}

\subsection{ Renormalization rule for the couplings }

The action of the operators  $\sigma^z_{2i-1}$ and $\sigma^z_{2i}$
on $\vert \lambda_{2i}^{S-} > $ of Eq. \ref{lambdasmeigen}
\begin{eqnarray}
\sigma^z_{2i-1} \vert \lambda_{2i}^{S-} > && =\sigma^z_{2i-1}
\frac{e^{  \frac{ \beta}{2} J^{(0)}_{2i-1,2i} } \vert u_S>+e^{ - \frac{ \beta}{2} J^{(0)}_{2i-1,2i} } \vert v_S>  }{\sqrt{ 2 \cosh \beta J^{(0)}_{2i-1,2i}} }
= \frac{e^{  \frac{ \beta}{2} J^{(0)}_{2i-1,2i} } \vert u_A>+e^{ - \frac{ \beta}{2} J^{(0)}_{2i-1,2i} } \vert v_A>  }{\sqrt{ 2 \cosh \beta J^{(0)}_{2i-1,2i}} }
\nonumber \\
\sigma^z_{2i} \vert \lambda_{2i}^{S-} > && =\sigma^z_{2i}
\frac{e^{  \frac{ \beta}{2} J^{(0)}_{2i-1,2i} } \vert u_S>+e^{ - \frac{ \beta}{2} J^{(0)}_{2i-1,2i} } \vert v_S>  }{\sqrt{ 2 \cosh \beta J^{(0)}_{2i-1,2i}} }
=
 \frac{e^{  \frac{ \beta}{2} J^{(0)}_{2i-1,2i} } \vert u_A>
- e^{ - \frac{ \beta}{2} J^{(0)}_{2i-1,2i} } \vert v_A>  }{\sqrt{ 2 \cosh \beta J^{(0)}_{2i-1,2i}} }
\label{opz2s}
\end{eqnarray}
and on $\vert \lambda_{2i}^{A-} > $ of Eq. \ref{eigenav}
\begin{eqnarray}
\sigma^z_{2i-1} \vert \lambda_{2i}^{A-}> && = \sigma^z_{2i-1}
\left[  u_{2i} \vert u_A> + v_{2i} \vert v_A> \right] =
u_{2i} \vert u_S> + v_{2i} \vert v_S>
\nonumber \\ 
\sigma^z_{2i} \vert \lambda_{2i}^{A-}> && = \sigma^z_{2i}
\left[  u_{2i} \vert u_A> + v_{2i} \vert v_A>\right]
= u_{2i} \vert u_S> - v_{2i} \vert v_S>
\label{opz2a}
\end{eqnarray}
leads to the projection rules
\begin{eqnarray}
P^-_{2i} \sigma^z_{2i-1} P^-_{2i} && = \mu_{2i-1} \sigma^z_{R(2i)}
\nonumber \\ 
P^-_{2i} \sigma^z_{2i} P^-_{2i}&& = \mu_{2i} \sigma^z_{R(2i)}
\label{progz12}
\end{eqnarray}
with the magnetic moments
\begin{eqnarray}
\mu_{2i-1} && = \frac{e^{  \frac{ \beta}{2} J^{(0)}_{2i-1,2i} } u_{2i} +e^{ - \frac{ \beta}{2} J^{(0)}_{2i-1,2i} } v_{2i}  }{\sqrt{ 2 \cosh \beta J^{(0)}_{2i-1,2i}} } 
\nonumber \\ 
\mu_{2i} && =
 \frac{e^{  \frac{ \beta}{2} J^{(0)}_{2i-1,2i} } u_{2i}
- e^{ - \frac{ \beta}{2} J^{(0)}_{2i-1,2i} } v_{2i}  }{\sqrt{ 2 \cosh \beta J^{(0)}_{2i-1,2i}} }
\label{rulesmu}
\end{eqnarray}
in terms of the coefficients $(u_i,v_i)$ introduced in Eq. \ref{UV}.

If two boxes $(2i-1,2i)$ and $(2j-1,2j)$ were initially coupled via the $k$-generation  couplings $(J^{(k)}_{2i-1,2j-1},J^{(k)}_{2i-1,2j},J^{(k)}_{2i,2j-1},J^{(k)}_{2i,2j})$,
the corresponding renormalized spins $\sigma_{R(2i)}$  and $\sigma_{R(2j)}$ 
will now be coupled via the renormalized coupling
\begin{eqnarray}
J^{R(k)}_{R(2i),R(2j)} && =  J^{(k)}_{2i-1,2j-1} \mu_{2i-1}\mu_{2j-1}
  + J^{(k)}_{2i-1,2j} \mu_{2i-1}\mu_{2j}
 +  J^{(k)}_{2i,2j-1}  \mu_{2i}\mu_{2j-1}
+ J^{(k)}_{2i,2j} \mu_{2i}\mu_{2j}
\label{jrj}
\end{eqnarray}

\section { Application to the Dyson pure ferromagnetic Ising model  }

\label{sec_ferro}

\subsection{ Model and notations }

In the Dyson pure ferromagnetic Ising model
\cite{dyson,bleher,gallavotti,book,jona,baker,mcguire,Kim,Kim77,dysonhopfield,us_rgdyndyson}, all couplings of a given generation $k$ 
in Eq. \ref{recDyson} coincide
\begin{eqnarray}
J^{(k)}_{i,j}= J^{(k)}
\label{jnpureferro}
\end{eqnarray}
and are chosen to decay exponentially with the generation number $n$
\begin{eqnarray}
J^{(k)} = \frac{J^{(0)}}{(2^k)^{1+\sigma}}=J^{(0)} 2^{-(1+\sigma)k}
\label{powerkferro}
\end{eqnarray}
in order to mimic the power-law behavior with respect to the distance $r=2^k$
\begin{eqnarray}
J(r) = \frac{J^{(0)}}{r^{1+\sigma}}
\label{powerr}
\end{eqnarray}
The parameter $\sigma$ is positive $\sigma>0$ in order to have an extensive energy.
Since the cost of a domain-wall at zero temperature scales as
\begin{eqnarray}
E^{DW}(L) \propto L^{1-\sigma} 
\label{edw}
\end{eqnarray}
the Dyson model mimics the non-hierarchical power-law ferromagnetic model only for
$0<\sigma \leq 1 $ where the energy cost of a Domain-Wall of Eq. \ref{edw} grows with the distance. The statics of this model has been much studied \cite{dyson,bleher,gallavotti,book,jona,baker,mcguire,Kim,Kim77}. In particular, the critical point between the ferromagnetic phase and the paramagnetic phase is mean-field-like for $0<\sigma \leq 1/2 $, and non-mean-field-like for
\begin{eqnarray}
\frac{1}{2}<\sigma \leq 1
\label{domain}
\end{eqnarray}
The real-space renormalization procedure described below
is expected to be appropriate in the non-mean-field-like region of Eq. \ref{domain}.

Here we are interested into the stochastic dynamics when
all the flipping times initially coincide
\begin{eqnarray}
\tau_i=\tau
\label{tauferroini}
\end{eqnarray}

\subsection{ First renormalization step }

The uniformity of the initial flipping times of Eq. \ref{tauferroini}
and the uniformity of the initial lowest order ferromagnetic coupling $J^{(0)}>0$
leads to many simplifications for the properties
 of the renormalized spins $\sigma_{R(2i)}$
describing boxes of two spins introduced in the previous section :

(i) they are all characterized by the same flipping time (Eq. \ref{tauR})
\begin{eqnarray}
\tau_{R} && = \tau e^{\beta  J^{(0)} } 
\label{tauRferro}
\end{eqnarray}

(ii) the coefficients of Eq. \ref{UV} simplify into
\begin{eqnarray}
u && = 1
\nonumber \\
v && = 0
\label{UVferro}
\end{eqnarray}
so that Eq. \ref{rulesmu} reduces to
\begin{eqnarray}
\mu_{2i-1} && = \mu_{2i} = \frac{e^{  \frac{ \beta}{2} J^{(0)} }  }
{\sqrt{ 2 \cosh \beta J^{(0)}} } =\frac{1  }{\sqrt{ 1+e^{- 2 \beta J^{(0)}}} } 
\label{rulesmuferro}
\end{eqnarray}
As a consequence, the RG rule of Eq. \ref{jrj} 
for the renormalized coupling becomes
\begin{eqnarray}
J^{R(k)}_{R(2i),R(2j)} && =  J^{(k)} \left[ \mu_{2i-1}+  \mu_{2i} \right]
\left[ \mu_{2j-1}+\mu_{2j} \right] =J^{(k)} \frac{ 4 }{ 1+e^{- 2 \beta J^{(0)}} } 
\label{jrjferro}
\end{eqnarray}
In particular, this renormalization of the couplings is
 independent of the renormalization of the flipping times of the dynamics.

\subsection{ RG flow of the couplings }

After one RG step, the initial couplings of Eq. \ref{powerk}
of the non-zero generations $k=1,2,..$ 
\begin{eqnarray}
J^{(k)} = J^{(0)} 2^{-(1+\sigma)k}
\label{powerk}
\end{eqnarray}
are renormalized into (Eq. \ref{jrjferro})
\begin{eqnarray}
J^{R(k)} &&  = 
 J^{(0)} \frac{ 2^{1-\sigma} }{ 1+e^{- 2 \beta J^{(0)}} } 2^{-(1+\sigma)(k-1)}
\label{jrjferroflowk}
\end{eqnarray}
It can be thus re-interpreted as the Dyson initial model
with generations $k-1=0,1,..$ with the lowest generation coupling
\begin{eqnarray}
J^{R(1)} &&  =
 J^{(0)} \frac{ 2^{1-\sigma} }{ 1+e^{- 2 \beta J^{(0)}} } 
\label{jrjferroflow}
\end{eqnarray}
The control parameter of the model is the ratio 
between the lowest generation coupling and the temperature :
the initial value
\begin{eqnarray}
K &&      \equiv  \beta  J^{(0)}
\label{defK}
\end{eqnarray}
is renormalized into
\begin{eqnarray}
K^R &&     \equiv \beta J^{R(1)} 
=  K    \frac{ 2^{1-\sigma} }{1+e^{  - 2 K } }  \equiv \Phi_{\sigma}(K)
\label{phidysonferro}
\end{eqnarray}
In the interesting region $0<\sigma<1$ (Eq \ref{domain}),
the RG flow $\Phi_{\sigma}(K) $ has two attractive fixed points :

(i) Near the ferromagnetic fixed point $K \to +\infty$,
 the corresponding linearized mapping
\begin{eqnarray}
K^R &&      \opsimeq_{K \to +\infty}  K  2^{1-\sigma}
\label{attractiveferro}
\end{eqnarray}
involves as it should the exponent
\begin{eqnarray}
\theta=1-\sigma
\label{thetadysonferro}
\end{eqnarray}
governing the energy cost of an interface (see Eq. \ref{edw}).

(ii) Near the paramagnetic fixed point $K \to 0$,
 the corresponding linearized mapping reads
\begin{eqnarray}
K^R &&      \opsimeq_{K \to 0}    2^{-\sigma} K
\label{attractivepara}
\end{eqnarray}

These two attractive fixed points are separated by the critical point 
 $K_c=\Phi_{\sigma}(K_c)$ describing the ferromagnetic/Paramagnetic transition.
 Its location as a function of the parameter $\sigma$
reads 
\begin{eqnarray}
 K_c(\sigma)  &&      = -\frac{1}{2}  \ln(  2^{1-\sigma}-1)
\label{critidysonferro}
\end{eqnarray}

The correlation length exponent $\nu(\sigma)$ 
 determined by the derivative of the RG flow
\begin{eqnarray}
2^{\frac{1}{\nu(\sigma)}} \equiv \Phi_{\sigma}'(K_c)  &&      =
1+ (2^{\sigma-1}-1) \ln(  2^{1-\sigma}-1)
\label{eqnudysonferro}
\end{eqnarray}
reads
\begin{eqnarray}
\nu(\sigma)  &&      = \frac{\ln 2}
{ \ln \left[ 1+ (2^{\sigma-1}-1) \ln(  2^{1-\sigma}-1) \right] }
\label{nudysonferro}
\end{eqnarray}

The magnetization exponent $x$ governing the finite-size power-law decay of
the magnetization at criticality
\begin{eqnarray}
m_L \propto L^{-x}
\label{defx}
\end{eqnarray}
is actually determined by the fixed point condition for the control parameter $K$
\begin{eqnarray}
K_c= \beta J_L L^{2(1-x)} = L^{-1-\sigma} \times L^{2-2x}= L^{1-\sigma-2x} 
\label{eqnudysonferrosdemi}
\end{eqnarray}
fixing
\begin{eqnarray}
x(\sigma) = \frac{1-\sigma}{2} 
\label{xdysonferro}
\end{eqnarray}
Equivalently, the two-point correlation scales as the square of the magnetization
\begin{eqnarray}
C(r)= m_r^2=  r^{-2x}= \frac{1}{r^{1-\sigma}}
\label{corre}
\end{eqnarray}
The identification with the standard form $1/r^{d-2+\eta}$ with $d=1$ yields that the exponent $\eta$ 
\begin{eqnarray}
\eta(\sigma)= 2-\sigma 
\label{etadysonferro}
\end{eqnarray}
keeps its mean-field value $(2-\sigma)$ even in the non-mean-field region
as it should \cite{luijtenblote2002}.

\subsection{ RG flow of the flipping times}

In terms of the control parameter $K$ of Eq. \ref{defK},
the first RG step of Eq. \ref{tauRferro} reads
\begin{eqnarray}
\tau^{R}  &&   =\tau   e^{\beta J^{(0)} } = \tau  e^{ K } 
\label{tauRdysonferro}
\end{eqnarray}
The iteration of this renormalization rule yields 
\begin{eqnarray}
\ln \frac{\tau^{R_n}}{\tau} &&   =  K+K^R+K^{R^2}+... + K^{R^{n-1}}
\label{tauRdysonferroiter}
\end{eqnarray}
in terms of the iterated values of the control parameter via the mapping
$\Phi_{\sigma}(K) $  of Eq. \ref{phidysonferro}.

\subsubsection{ Dynamical Exponent at the critical point $K_c$ }

At the critical fixed point $K_c$ of Eq. \ref{critidysonferro}, 
Eq. \ref{tauRdysonferroiter}
reduces to
\begin{eqnarray}
\ln \frac{\tau (L=2^n)}{\tau} &&   = n K_c = \frac{\ln L}{\ln 2} K_c
\label{tauRdysonferron}
\end{eqnarray}
so that the characteristic time $\tau(L)$
associated to the spatial length $L$ follows the critical power-law scaling
\begin{eqnarray}
\tau(L) \propto L^{z_c}
\label{defzcferro}
\end{eqnarray}
with the dynamical exponent
\begin{eqnarray}
z_c(\sigma) = \frac{ K_c}{\ln 2} =-\frac{1}{2 \ln 2 }  \ln(  2^{1-\sigma}-1)
\label{zcdysonferro}
\end{eqnarray}

\subsubsection{ Barrier exponent in the ferromagnetic phase  }

Near the ferromagnetic fixed point $K \to +\infty$
 with the linearized mapping of Eq. \ref{attractiveferro},
Eq. \ref{tauRdysonferroiter} leads to
 the activated scaling
\begin{eqnarray}
\ln \frac{\tau(L=2^n)}{\tau} && 
  \simeq K \sum_{k=0}^{n-1} 2^{(1-\sigma)k} = K \frac{2^{(1-\sigma)n}-1}
{2^{(1-\sigma)}-1} =  K \frac{L^{(1-\sigma)}-1}{2^{(1-\sigma)}-1} \propto L^{\psi}
\label{tauRdysonferroactivated}
\end{eqnarray}
with the barrier exponent
\begin{eqnarray}
\psi = 1-\sigma = \theta
\label{psi}
\end{eqnarray}
This activated scaling of the largest relaxation time is in agreement with
the previous study concerning the very-low temperature \cite{us_rgdyndyson}.

\subsubsection{ Relaxation in the in the paramagnetic phase  }

Near the paramagnetic fixed point $K \to 0$ with the linearized mapping of Eq. \ref{attractivepara}, one obtains that the renormalized flipping time
\begin{eqnarray}
\ln \frac{\tau(L=2^n)}{\tau} &&   \simeq K \sum_{k=0}^{n-1} 2^{-\sigma k} = K \frac{1- 2^{-\sigma n}}
{1-2^{-\sigma} } 
\label{tauRdysonferropara}
\end{eqnarray}
remains finite as it should.

\section { Application to the Dyson Spin-Glass model  }

\label{sec_sg}

\subsection{ Model and notations }

 In the Dyson hierarchical spin-glass model
 \cite{franz,castel_etal,castel_parisi,castel,angelini,guerra,barbieri,us_LRSGground,us_LRSGfull}, the couplings $J^{k}(i,j) $ 
of generation $k$, associated to the length scale $L_k=2^k$ read
\begin{eqnarray}
J_k(i,j)=\Delta_k \epsilon_{ij}
\label{jndysonsg}
\end{eqnarray}
where the $ \epsilon_{ij}$ are independent random variables of zero mean
 distributed with the Gaussian law
 \begin{eqnarray}
G(\epsilon)  =  \frac{1}{\sqrt{4 \pi} } e^{- \frac{\epsilon^2}{4}}
\label{gaussian}
\end{eqnarray}
The characteristic scale $\Delta_k$ is chosen to decay exponentially
with the number $k$ of generations
\begin{eqnarray}
\Delta_k = 2^{-k \sigma} =  \frac{1}{L_k^{\sigma}} 
\label{deltandysonsg}
\end{eqnarray}
in order to mimic the power-law decay 
\begin{eqnarray}
\Delta(L) && = \frac{1}{L^{\sigma}}
\label{defsigma}
\end{eqnarray}
of the much studied non-hierarchical Long-Ranged Spin-Glass chain
 \cite{kotliar,BMY,Fis_Hus,KY,KYgeom,KKLH,KKLJH,Katz,KYalmeida,Yalmeida,KDYalmeida,LRmoore,KHY,KH,mori,wittmann,us_overlaptyp,us_conjugateLRSG,us_chaos,us_LRSGground,us_LRSGfull,wittmann_dyn,alain}.

The energy is extensive in the region
\begin{eqnarray}
 \sigma  > \frac{1}{2}
\label{regionmu}
\end{eqnarray}
The energy cost of an interface at zero temperature
 \begin{eqnarray}
E^{DW}_L \propto L^{\theta}
\label{deftheta}
\end{eqnarray}
involves the droplet exponent predicted via scaling arguments \cite{Fis_Hus,BMY}
\begin{eqnarray}
\theta^{LR}_{Gauss}(d=1,\sigma) && = 1-\sigma 
\label{thetaLRd1gauss}
\end{eqnarray}
The Dyson hierarchical model thus mimics 
correctly the non-hierarchical long-ranged spin-glass
model when the droplet exponent is positive $\theta^{LR}_{Gauss}(d=1,\sigma)  = 1-\sigma >0 $. Then the critical point between
 the spin-glass phase and the paramagnetric phase is
expected to be mean-field-like for $1/2<\sigma<2/3$, 
and non-mean-field-like for \cite{kotliar}
\begin{eqnarray}
\frac{2}{3} < \sigma <1
\label{regionsg}
\end{eqnarray}
The real-space renormalization procedure described below
is expected to be appropriate in the non-mean-field-like region of Eq. \ref{regionsg}.

Here even if the flipping times coincide initially
\begin{eqnarray}
\tau_i=\tau
\label{tauferro}
\end{eqnarray}
 they will not remain uniform upon renormalization, 
in contrast to the ferromagnetic case discussed in the previous section. 
As a consequence, the renormalization of the couplings of Eq. \ref{jrj}
is not independent of the renormalization of the flipping times.
Physically, this means that for the spin-glass at finite temperature,
the renormalization for the statics as described by the couplings
cannot be formulated in closed form independently of the dynamical properties
that determine the appropriate renormalized spins, i.e.  the appropriate valleys separated by
the biggest flipping times.

Let us first describe the limits near zero or infinite temperature
before presenting the numerical results at finite temperature.

\subsection{ RG flow near zero temperature $T \to 0$ }

In the limit of small temperature $T \to 0$ i.e. $\beta=1/T \to +\infty$,
the coefficients of Eq. \ref{UV} simplify into
\begin{eqnarray}
u_{2i} && \simeq \delta_{{\rm sgn }J^{(0)}_{2i-1,2i} ,+1} 
\nonumber \\
v_{2i} && \simeq \delta_{{\rm sgn} J^{(0)}_{2i-1,2i} ,-1}
\label{UVzerotemp}
\end{eqnarray}
and the coefficients of Eq. \ref{rulesmu} into
\begin{eqnarray}
\mu_{2i-1} && \simeq 1
\nonumber \\ 
\mu_{2i} && \simeq {\rm sgn} J^{(0)}_{2i-1,2i}
\label{rulesmuzerotemp}
\end{eqnarray}
so that the RG rule of Eq. \ref{jrj} reduces to
\begin{eqnarray}
J^{R(k)}_{R(2i),R(2j)} && =  J^{(k)}_{2i-1,2j-1} 
  + J^{(k)}_{2i-1,2j} {\rm sgn} J^{(0)}_{2j-1,2j}
 +  J^{(k)}_{2i,2j-1}  {\rm sgn} J^{(0)}_{2i-1,2i}
+ J^{(k)}_{2i,2j} {\rm sgn} J^{(0)}_{2i-1,2i} {\rm sgn} J^{(0)}_{2j-1,2j}
\label{jrjsgzerot}
\end{eqnarray}
This rules coincides with the zero-temperature RG studied in detail in
\cite{us_LRSGground} and its growth is governed by the droplet 
exponent $\theta=1-\sigma$ of Eq. \ref{thetaLRd1gauss} as it should
(see \cite{us_LRSGground} for more details)
 \begin{eqnarray}
J_L \propto L^{\theta}
\label{flowtheta}
\end{eqnarray}

In the limit of small temperature $T \to 0$ i.e. $\beta=1/T \to +\infty$,
the renormalized flipping time of Eq. \ref{tauR} follows the Arrhenius behavior 
\begin{eqnarray}
\tau_{R(2i)} 
&& \opsimeq_{\beta \to +\infty} \frac{
  \left(\tau_{2i-1} +\tau_{2i}  \right)   } { 2 }e^{ \beta \vert J^{(0)}_{2i-1,2i} \vert}
\label{tauRzerot}
\end{eqnarray}
and is similar to the low-temperature renormalization of the Metropolis dynamics
discussed in detail in \cite{us_LRSGfull} : the barrier exponent 
$\psi$ characterizing the activated dynamics
\begin{eqnarray}
\ln \tau(L) \propto L^{\psi}
\label{tauRzerotactivated}
\end{eqnarray}
is found to coincide with the droplet exponent (see \cite{us_LRSGfull}
for more details)
 \begin{eqnarray}
\psi=\theta=1-\sigma
\label{psithetasg}
\end{eqnarray}

\subsection{ RG flow at high temperature $T \to +\infty$ }

In the limit of high temperature $T \to +\infty$ i.e. $\beta \to 0$,
the renormalized flipping time of Eq. \ref{tauR} reduces to
\begin{eqnarray}
\tau_{R(2i)}
&& \opsimeq_{\beta \to 0} \frac{
  \left(\tau_{2i-1} +\tau_{2i}  \right)  
+ \vert \tau_{2i-1}   - \tau_{2i}  \vert  } { 2 } = {\rm max } (\tau_{2i-1},\tau_{2i})
\label{tauRinft}
\end{eqnarray}
so that the flipping times remain finite.

\begin{figure}[htbp]
 \includegraphics[height=6cm]{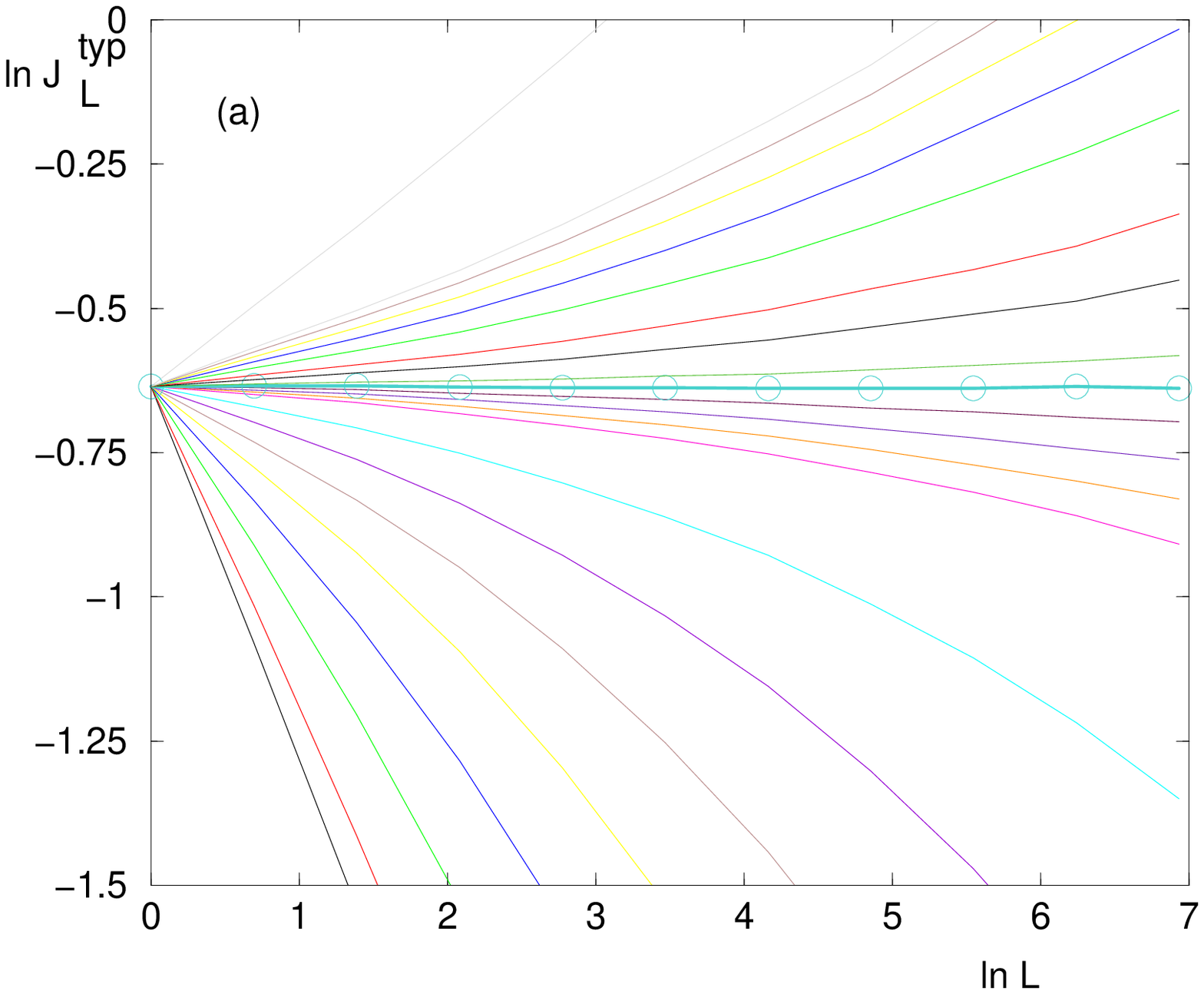}
\hspace{1cm}
 \includegraphics[height=6cm]{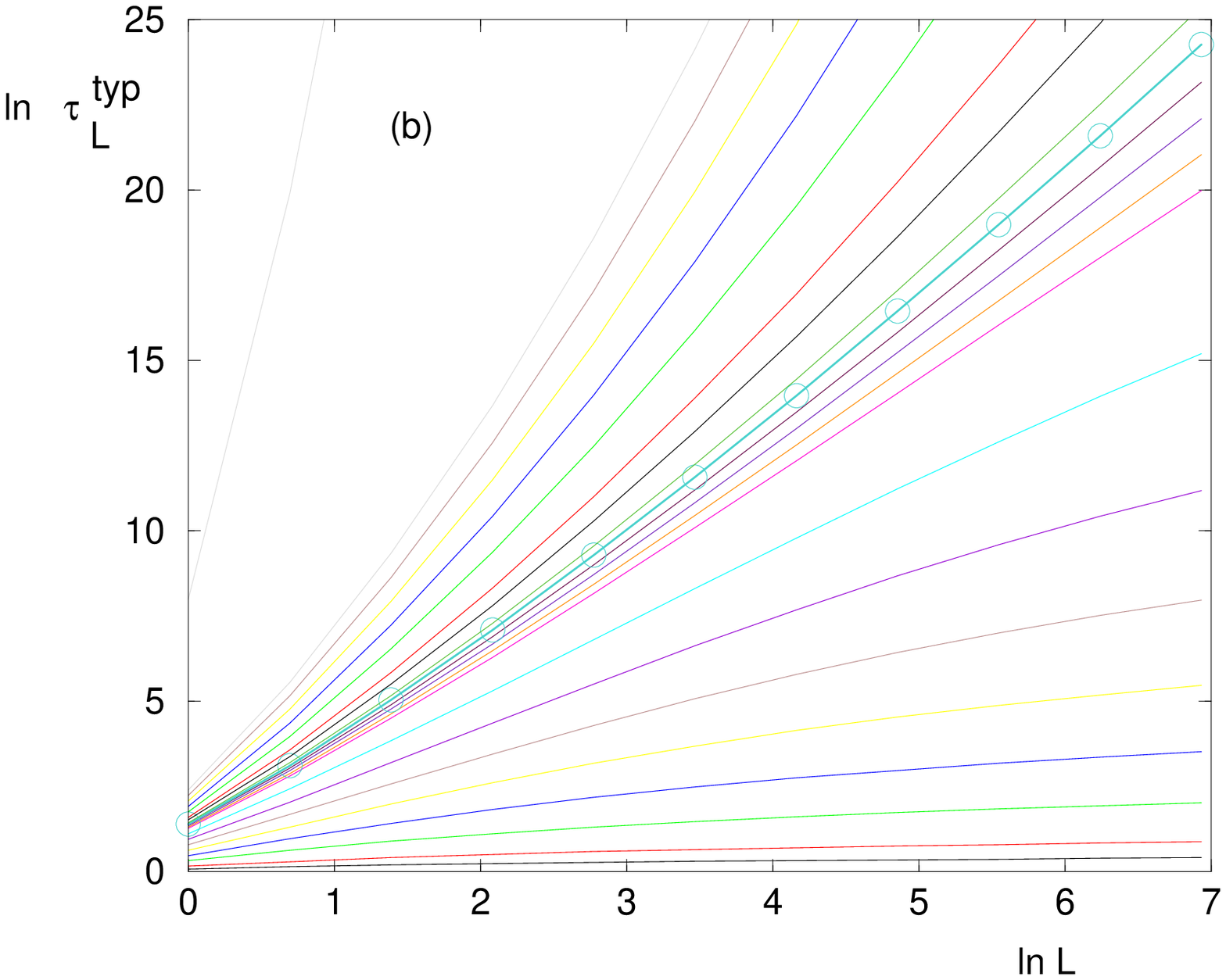}
\caption{ Numerical application of the RG rules for the
Dyson Spin-Glass model of parameter $\sigma=0.75$ \\
(a) RG flow  of the logarithm of the typical coupling 
$\ln J_L^{typ} \equiv \overline{ \ln \vert J_L \vert}$  as a function of $\ln L$ : \\
the growth in the spin-glass phase $\beta>\beta_c$ and the decay in the paramagnetic phase $\beta<\beta_c$ are separated by the critical unstable fixed point $\beta_c \simeq 1.76$
(circles)  \\
(b) RG flow of the typical relaxation time 
$\ln \tau_L^{typ} \equiv \overline{ \ln \tau_L }$  as a function of $\ln L$ :
the slope at the critical point $\beta_c \simeq 1.76$ corresponds to the dynamical exponent $z_c \simeq 3.75$ (circles); in the paramagnetic phase $\beta<\beta_c$, the relaxation time converges towards a constant; in the spin-glass phase $\beta>\beta_c$, the growth corresponds to the activated scaling with the droplet exponent (see Eq. \ref{tauRzerotactivated}).  }
\label{figflow}
\end{figure}

\subsection{ Numerical study of the RG flows in the critical region }

For the Dyson chain of parameter $\sigma=0.75$,
the RG rules have been numerically applied to $n_s=13.10^3$ disordered samples
containing $n=12$ generations corresponding to $N=2^{12} =4096$ spins.
On Fig. \ref{figflow} (a) is shown the RG flow of the logarithm of the typical coupling
\begin{eqnarray}
\ln J_L^{typ}  \equiv \overline{ \ln \vert J_L \vert}
\label{jtypflow}
\end{eqnarray}
The critical point corresponds to the unstable fixed point between the growth
$\ln J_L^{typ} \propto \theta \ln L = (1-\sigma) \ln L  $ of the Spin-Glass phase
and the decay of the Paramagnetic phase :
the inverse critical temperature is around $\beta_c \simeq 1.76$.

On Fig. \ref{figflow} (b) is shown the RG flow of the logarithm of the typical relaxation time
\begin{eqnarray}
\ln \tau_L^{typ}  \equiv \overline{ \ln  \tau_L }
\label{tautypflow}
\end{eqnarray}
The critical point $\beta_c \simeq 1.76$ corresponds to the power-law scaling
with a dynamical exponent of order $z_c \simeq 3.75 $,
whereas the Spin-Glass phase is characterized by the activated scaling of Eq \ref{tauRzerotactivated}, and the Paramagnetic phase is characterized by a finite relaxation time.

\section{ Conclusion }

\label{sec_conclusion}

In this article, we have proposed to consider the Boltzmann equilibrium as the stationary measure of the stochastic dynamics satisfying detailed balance and to study via real-space renormalization the corresponding master equation. This point of view has the following advantages :

(i) for the case of complex systems with frustration like spin-glasses, 
one obtains that the renormalization rules 
for the spatial couplings depend upon the relaxation times of the dynamics. 
This suggests that at finite temperature $T$,
it is not possible to write consistent closed renormalization rules for the couplings alone, in contrast to the zero temperature limit where it is possible \cite{us_LRSGground}.

(ii) for the case of unfrustrated models like ferromagnets, 
one obtains that the renormalization rules 
for the spatial couplings are independent of the dynamical variables. 
However even in this case, the dynamical framework
to derive the static RG rules is useful to select the two appropriate states defining the renormalized spins and to avoid the arbitrariness of other definitions of renormalized spins that are used in other real-space RG schemes (such as the majority rule for instance).

Focusing on Dyson hierarchical spin models, we have derived 
the explicit RG rules for the couplings and for the relaxation times.
For the ferromagnetic case, the RG flows have been explicitly solved,
while for the spin-glass case, the RG rules have been numerically applied in
the critical region. In both cases, we have obtained that the characteristic relaxation time remains finite in the high-temperature paramagnetic phase, 
follows some activated dynamics in the low-temperature phase (either ferromagnetic or spin-glass), and displays
some critical power-law scaling with some dynamical exponent $z_c$
at the phase transition.

\end{document}